\def\year{2022}\relax
\documentclass[letterpaper]{article} 
\usepackage{aaai22}  
\usepackage{times}  
\usepackage{helvet}  
\usepackage{courier}  
\usepackage[hyphens]{url}  
\usepackage{graphicx} 
\usepackage{natbib}  
\usepackage{caption} 
\DeclareCaptionStyle{ruled}{labelfont=normalfont,labelsep=colon,strut=off} 
\usepackage{algorithm}
\usepackage{algorithmic}

\usepackage{newfloat}
\usepackage{listings}
\floatstyle{ruled}
\newfloat{listing}{tb}{lst}{}
\floatname{listing}{Listing}

\usepackage{amssymb,amsmath,amsthm,enumerate,threeparttable,bm,graphicx,subfigure}
\usepackage{algorithm, algorithmic}
\usepackage{booktabs}
\usepackage{multirow}
\usepackage{lipsum}
\usepackage{enumitem}
\usepackage{xcolor}  
\usepackage{appendix}

\newtheorem{definition}{Definition}

\long\def\comment#1{}
\usepackage{setspace}
\usepackage{amsmath}
\usepackage{wrapfig}
\usepackage[export]{adjustbox}
\usepackage[switch]{lineno}

\usepackage{array}
\newcommand{\tabincell}[2]{\begin{tabular}{@{}c#1@{}}#2\end{tabular}} 

\title{Defending against Model Stealing via Verifying Embedded External Features}
\author {
    Yiming Li$^{1,}$\thanks{The first two authors contributed equally to this work.}, Linghui Zhu$^{1,2,*}$, Xiaojun Jia$^{3}$, Yong Jiang$^{1,2}$, Shu-Tao Xia$^{1,2,}$\thanks{Corresponding Author: Shu-Tao Xia}, Xiaochun Cao$^{3}$
}
\affiliations{
    $^1$Tsinghua Shenzhen International Graduate School, Tsinghua University, Shenzhen, China\\
    $^2$Peng Cheng Laboratory, Shenzhen, China\\
    $^3$Institute of Information Engineering, Chinese Academy of Sciences, Beijing, China\\
    li-ym18@mails.tsinghua.edu.cn; xiast@sz.tsinghua.edu.cn
}

\begin{document}

\maketitle

\begin{abstract}
Obtaining a well-trained model involves expensive data collection and training procedures, therefore the model is a valuable intellectual property. Recent studies revealed that adversaries can `steal' deployed models even when they have no training samples and can not get access to the model parameters or structures. Currently, there were some defense methods to alleviate this threat, mostly by increasing the cost of model stealing. In this paper, we explore the defense from another angle by verifying whether a suspicious model contains the knowledge of defender-specified \emph{external features}. Specifically, we embed the external features by tempering a few training samples with style transfer. We then train a meta-classifier to determine whether a model is stolen from the victim. This approach is inspired by the understanding that the stolen models should contain the knowledge of features learned by the victim model. We examine our method on both CIFAR-10 and ImageNet datasets. Experimental results demonstrate that our method is effective in detecting different types of model stealing simultaneously, even if the stolen model is obtained via a multi-stage stealing process. The codes for reproducing main results are available at Github (https://github.com/zlh-thu/StealingVerification).
\end{abstract}

\section{Introduction}
Deep learning, especially deep neural networks (DNNs), has demonstrated its great power in many applications \cite{guo2020deep,stokes2020deep,minaee2021image}.
In general, training a well-performed model requires a large number of training samples and a massive amount of computational resources. Both data collection and training process are expensive and time-consuming, which makes the trained model a valuable intellectual property to its owner.

Recently, researchers found that adversaries can `steal' the (deployed) \emph{victim model} even when they have no training samples and can not get access to the model parameters or structures \cite{tramer2016stealing,orekondy2019knockoff,chandrasekaran2020exploring}. For example, the adversaries may use the victim model to label an unlabeled dataset, based on which to train the \emph{stolen model}. This threat is called \emph{model stealing}. Since the model stealing can obtain a function-similar copy of the victim model stealthily, it poses a huge threat to model owners.

\begin{figure}[!t]
    \centering
    \includegraphics[width=0.43\textwidth]{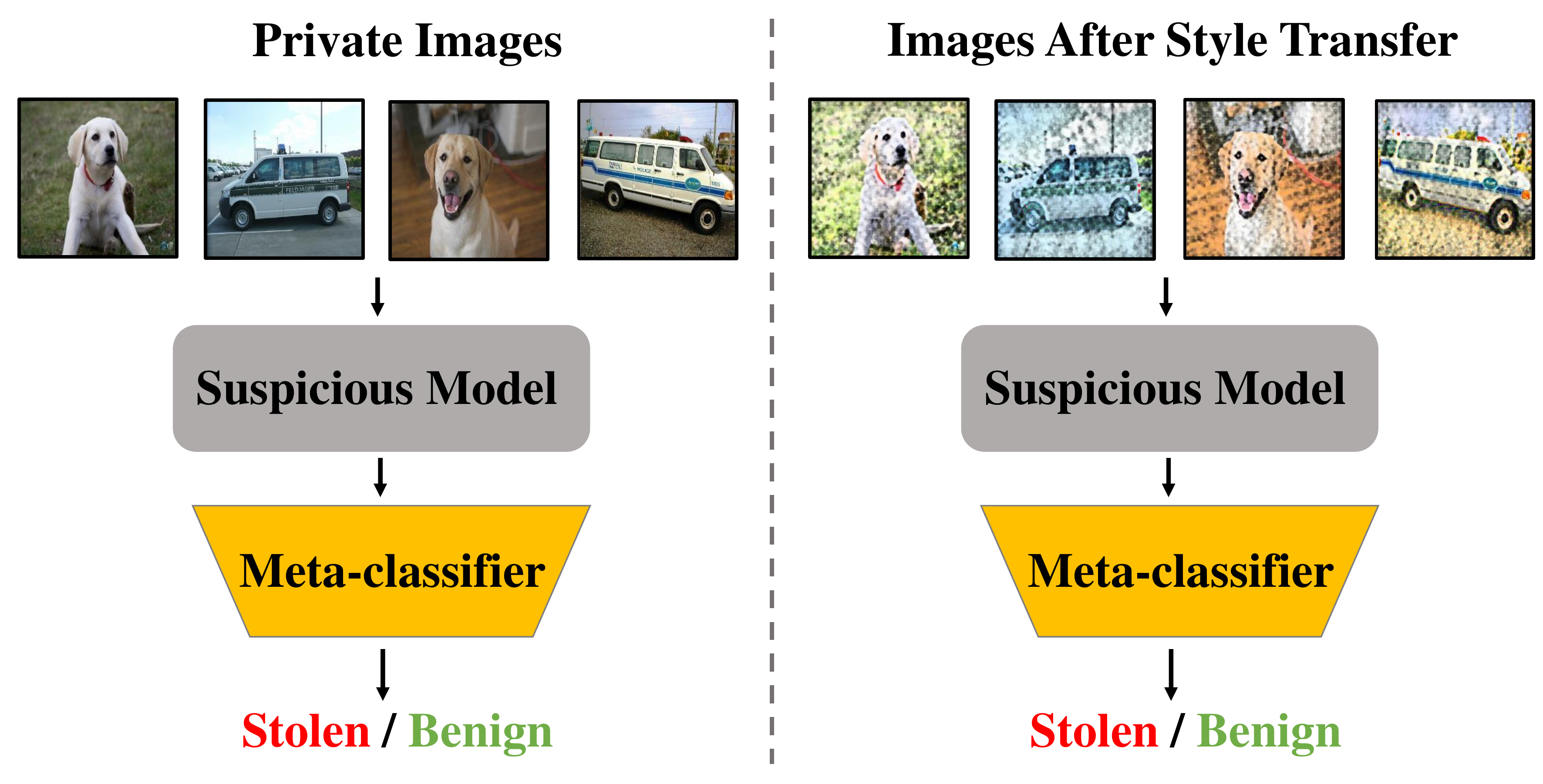}
    \vspace{-0.8em}
    \caption{The verification stage of dataset inference and our method. Specifically, dataset inference adopts inherent features contained in benign private images while our method utilizes external features embedded in stylized images.}
    \label{fig:intro}
    \vspace{-0.8em}
\end{figure}

To alleviate the threat of model stealing, there were also some defense methods, mostly by introducing randomness or perturbation in the victim models to increase the costs of model stealing \cite{tramer2016stealing,lee2018defending,kariyappa2020defending}. For instance, defenders may perturb the prediction by rounding or adding noise to the posterior probabilities. However, these defenses may significantly reduce the performance of legitimate users and could even be bypassed by following adaptive attacks \cite{jia2021entangled,maini2021dataset}.

In this paper, we explore the defense of model stealing from another perspective by \emph{verifying whether a suspicious model has defender-specified behaviors}. If the model has such behaviors, we treat it as stolen from the victim. This approach is inspired by the understanding that the stolen models should contain the knowledge of features learned by the victim model and therefore they have similar behaviors. To the best of our knowledge, there is only one work, $i.e.$, \emph{dataset inference} \cite{maini2021dataset}, focusing on this perspective, where they adopted \emph{inherent features} of the training set to verify model ownership. However, we reveal that this approach is easy to make misjudgments, especially when the training set of suspicious models have a similar distribution to that of the victim model. The misjudgment is most probably because different models may learn similar inherent features once their training sets have certain similarities. Based on this understanding, we propose to embed defender-specified \emph{external features} into victim models for ownership verification. These features are different from those contained in the original training set. Specifically, we embed external features by tempering the images of some training samples with \emph{style transfer}. Since we only poison a few samples and do not change their labels, the embedded features will not hinder the functionality of the victim model. Besides, we also train a \emph{benign model} based on the original training set. It is used only for training a \emph{meta classifier} to determine whether a suspicious model is stolen from the victim model. Only the model containing the knowledge of external features will be deployed.

\vspace{0.3em}
The main contribution of this work is four-fold: \textbf{(1)} We revisit the defense of model stealing from the aspect of ownership verification. \textbf{(2)} We reveal the limitations of existing verification-based methods. Based on these understandings, we propose a simple yet effective defense approach. \textbf{(3)} We verify the effectiveness of our method on benchmark datasets under various types of model stealing simultaneously. \textbf{(4)} Our work could provide a new angle about how to adopt `data poisoning' for positive purposes.

\vspace{0.4em}
\section{Background and Related Work}
In this paper, we focus on model stealing and its defenses towards image classification. Other tasks are out the scope of this paper. We will discuss them in our future work.


\subsection{Model Stealing}
\label{sec:stealing-attack}
Model stealing aims to steal the intellectual property from a victim by obtaining a function-similar copy of the deployed model. 
In general, existing methods can be divided into three categories based on adversary's permission level, as follows:

\vspace{0.3em}
\noindent \textbf{Dataset-Accessible Attacks ($\mathcal{A}_D$): }
In this setting, the adversary can get access to the training dataset whereas can only query the model. In this case, the adversary can train a substitute model based on knowledge distillation \cite{hinton2015distilling} for model stealing.

\vspace{0.3em}
\noindent \textbf{Model-Accessible Attacks ($\mathcal{A}_M$): }
In this setting, the adversary has complete access to the victim model. This type of attack could happen when the victim model is open-sourced or via insider access. In this case, the adversary can obtain a substitute model by using data-free knowledge distillation based on zero-shot learning \cite{fang2019data} or simply by fine-tuning the victim model with local training samples.

\vspace{0.3em}
\noindent \textbf{Query-Only Attacks ($\mathcal{A}_Q$): } 
In this setting, the adversary can only query the model. Query-only attacks can also be divided into two subclasses, including the \emph{label-query attacks} \cite{papernot2017practical,jagielski2020high,chandrasekaran2020exploring} and \emph{logit-query attacks} \cite{tramer2016stealing,orekondy2019knockoff}, based on model's feedback. In general, label-query attacks adopted the victim model to annotate some substitute (unlabeled) samples, based on which to train their substitute model. In the logit-query attacks, the adversary usually obtains the substitute model by minimizing the distance between its predicted logits and those generated by the victim model.

\subsection{Defenses against Model Stealing}
\label{sec:defense}
\noindent \textbf{Non-verification based Defenses. }
Currently, most of the existing methods alleviated the stealing threat by increasing the cost of model stealing through perturbing model results. For instance, defenders could round the probabilities \cite{tramer2016stealing}, added noise to the prediction that results in a high loss \cite{lee2018defending}, only returning the most confident label \cite{orekondy2019knockoff}. However, these defenses may significantly reduce the performance of legitimate users and could even be bypassed by adaptive attacks \cite{jia2021entangled,maini2021dataset}. Other works \cite{kesarwani2018model,juuti2019prada, yan2021monitoring} detected model stealing by identifying malicious queries. However, these methods relied on some assumptions of malicious query patterns, which may not be adopted by the adversaries in practice.

\vspace{0.3em}
\noindent \textbf{Dataset Inference. }
To the best of our knowledge, this is the first and currently the only verification-based defense against model stealing. Its key idea is to identify whether a suspicious model contains the knowledge of the inherent features that the victim model $V$ learned from the private training set. Specifically, let we consider a $K$-classification problem. For each sample $(\bm{x}, y)$, dataset inference first generated its minimum distance $\bm{\delta}_t$ to each class $t$ by 
\begin{equation}
    \min_{\bm{\delta}_t} d(\bm{x}, \bm{x}+\bm{\delta}_t), s.t., V(\bm{x}+\bm{\delta}_t) = t,
\end{equation}
where $d(\cdot)$ is a distance metric ($e.g.$, $\ell^\infty$ norm). The distance to each class $\bm{\delta}=(\bm{\delta}_1, \cdots, \bm{\delta}_K)$ is the feature embedding of sample $(\bm{x}, y)$ $w.r.t.$ the victim model $V$. After that, the defender will randomly select some samples inside (labeled as `+1') or out-side (labeled as `-1') their private dataset and use the feature embedding $\bm{\delta}$ to train a binary meta-classifier $C$, where $C(\bm{\delta}) \in [0,1]$ indicates the probability that the sample $(\bm{x}, y)$ is from the private set. To determine whether a suspicious model is stolen from the victim, the defender creates equal-sized sample vectors from private and public samples and conduct the \emph{hypothesis test} based on the trained $C$. If the confidence scores of private samples are significantly greater than those of public samples, the suspicious model is treated as stolen from the victim. However, as shown in following experiments, this method is easy to make misjudgments, especially when the training set of suspicious models have a similar distribution to that of the victim model. 
This limitation hinders its utility in practice.

\vspace{0.5em}
\noindent \textbf{Model Watermarking. }
The main purpose of model watermarking is detecting theft ($i.e.$, directly copy the model) instead of preventing model stealing. However, we notice that the dataset inference enjoys certain similarities to the \emph{misclassification-based model watermarking} \cite{adi2018turning,li2020open,jia2021entangled}, especially the backdoor-based ones \cite{adi2018turning,zhang2018protecting,li2020open}. As such, these approaches could be potential defenses against model stealing. Specifically, these methods performed ownership verification by making the protected model misclassifying defender-specified samples. For example, defenders may first adopt \emph{backdoor attacks} \cite{gu2019badnets,li2020backdoor,nguyen2020input,li2021backdoor,bagdasaryan2021blind,li2021invisible} to watermark the model during the training process and then conduct the ownership verification. In general, a backdoor attack can be characterized by three components, including the trigger pattern $\bm{t}$, target class $y_t$, and adversary-predefined poisoned image generator $G(\cdot)$. Given the benign training set $\mathcal{D} = \{ (\bm{x}_i, y_i) \}_{i=1}^{N}$, the backdoor adversary will randomly select $\gamma \%$ samples ($i.e.$, $\mathcal{D}_s$) from $\mathcal{D}$ to generate their poisoned version $\mathcal{D}_p = \{ (\bm{x}', y_t) |\bm{x}' = G(\bm{x}; \bm{t}), (\bm{x}, y) \in \mathcal{D}_s \}$. Different backdoor attacks may assign different generator $G(\cdot)$. For example, $G(\bm{x}; \bm{t}) = (\bm{1}-\bm{\lambda}) \otimes \bm{x}+ \bm{\lambda} \otimes \bm{t}$ where $\bm{\lambda} \in \{0,1\}^{C \times W \times H}$ and $\otimes$ indicates the element-wise product in the BadNets \cite{gu2019badnets}. 
After $\mathcal{D}_p$ was generated, $\mathcal{D}_p$ and remaining benign samples $\mathcal{D}_b \triangleq \mathcal{D} \backslash \mathcal{D}_s$ will be used to train the model $f_\theta$ via

\begin{equation}
\min_{\bm{\theta}} \sum_{(\bm{x}, y) \in \mathcal{D}_p \cup \mathcal{D}_b} \mathcal{L}(f_{\bm{\theta}}(\bm{x}), y).
\end{equation}
In the verification stage, the defender will examine suspicious models in predicting $y_t$. If the confidence scores of poisoned samples are significantly greater than those of benign samples, the suspicious model is treated as watermarked and therefore it is stolen from the victim. However, as shown in the following experiments, these methods have far less effective in defending against model stealing. 

\vspace{0.3em}
\section{Revisiting Verification-based Defenses}
\label{sec:limitation}

\vspace{0.3em}
\subsection{The Limitation of Dataset Inference}
\label{sec:lim_of_inherent}

As illustrated in Section \ref{sec:defense}, dataset inference relied on a latent assumption that a model will not learn the features contained in the private dataset if it is not stolen from the victim. However, since different models may learn similar features even they are trained on different datasets, this assumption does not hold and therefore the method may misjudge. In this section, we verify this limitation.

\vspace{0.3em}
\noindent \textbf{Settings. }
In this section, we conduct the experiments on CIFAR-10 \cite{krizhevsky2009learning} dataset with VGG \cite{simonyan2014very} and ResNet \cite{he2016deep}. Specifically, we randomly separate the original training set $\mathcal{D}$ into two disjoint subsets $\mathcal{D}_l$ and $\mathcal{D}_r$. We train the VGG on $\mathcal{D}_l$ (dubbed VGG-$\mathcal{D}_l$) and the ResNet on $\mathcal{D}_r$ (dubbed ResNet-$\mathcal{D}_r$), respectively. We also train the VGG on a noisy dataset $\mathcal{D}_l' \triangleq \{(\bm{x}', y)|\bm{x}' = \bm{x} + \mathcal{N}(0, 16), (\bm{x}, y) \in \mathcal{D}_l\}$ (dubbed VGG-$\mathcal{D}_l'$) for reference. In the verification process, we verify whether the VGG-$\mathcal{D}_l$ and VGG-$\mathcal{D}_l'$ is stolen from ResNet-$\mathcal{D}_r$ and whether the ResNet-$\mathcal{D}_r$ is stolen from VGG-$\mathcal{D}_l$ based on the settings proposed in dataset inference \cite{maini2021dataset}. Besides, we also adopt the p-value as the evaluation metric. The p-value is calculated based on the approach described in Section \ref{sec:defense}. Note that \emph{the smaller the p-value, the more confident that dataset inference believes the model stealing happened}. More detailed settings are in \textbf{Appendix}.

\begin{table}[t]
\centering
\scalebox{0.97}{
\begin{tabular}{c|ccc}
\toprule  
& ResNet-$\mathcal{D}_r$ & VGG-$\mathcal{D}_l$ & VGG-$\mathcal{D}_l'$ \\ \hline
Accuracy & 88.0\% & 87.7\% & 85.0\% \\
p-value  & $\bm{10^{-7}}$ & $\bm{10^{-5}}$  & $\bm{10^{-4}}$ \\ \bottomrule
\end{tabular}
}
\vspace{-0.8em}
\caption{The accuracy of victim models and p-value of verification processes. Dataset inference misjudged in all cases. }
\label{table:misjudge}
\vspace{-0.2em}
\end{table}

\begin{table}[ht]
\centering
\scalebox{0.95}{
\begin{tabular}{c|ccc}
\toprule  
Model Type $\rightarrow$  & Benign & Watermarked & Stolen \\ \hline
BA                        & 91.99& 90.03    & 70.17 \\\hline
ASR                       & 0.01 & 98.02   & $\bm{3.84}$ \\ \bottomrule
\end{tabular}
}
\vspace{-0.8em}
\caption{The performance (\%) of different models.}
\label{table:models}
\vspace{-0.5em}
\end{table}

\vspace{0.35em}
\noindent \textbf{Results. }
As shown in Table \ref{table:misjudge}, all models achieve decent performance even when the training samples are limited.
However, the p-value is significantly smaller than 0.01 in all cases. In other words, the dataset inference believes that these models are stolen from the victim in a high confidence. However, in each case, since the suspicious and the victim model are trained on completely different training samples and with different model structures, the suspicious model should not be considered as stolen from the victim. These results reveal that \emph{the dataset inference could make misjudgments and therefore its results are questionable}. In particular, the p-value of VGG-$\mathcal{D}_l$ is smaller than that of the VGG-$\mathcal{D}_l'$. This is probably because the latent distribution of $\mathcal{D}_l'$ is more different from that of $\mathcal{D}_r$ (compared with that of $\mathcal{D}_l$) and therefore models learn more different features.

\subsection{The Limitation of Model Watermarking}
\label{sec:lim_of_external}
Intuitively, the inference process of backdoor attacks is similar to unlocking a door with the corresponding key. As such, the success of backdoor-based model watermarking relied on an assumption that the trigger pattern matches hidden backdoors contained in the stolen model. This assumption holds in its originally discussed scenarios where the stolen model is the same as the victim model. However, it may not hold in the model stealing, since the backdoors contained in the stolen models may be changed or even removed during the stealing process. Accordingly, backdoor-based model watermarking may fail in defending against model stealing. In this section, we verify this limitation.

\vspace{0.3em}
\noindent \textbf{Settings. }
In this part, we adopt the most representative and effective backdoor attack, the BadNets \cite{gu2019badnets}, as an example for the discussion. The watermarked model will then be stolen by the data-free distillation-based model stealing \cite{fang2019data}. We adopt the \emph{benign accuracy (BA)} and \emph{attack success rate (ASR)} \cite{li2020backdoor} to evaluate the performance of the stolen model. The larger the ASR, the more likely the stealing will be detected. More detailed settings can be found in \textbf{Appendix}.

\begin{figure*}[ht]
    \centering
    \includegraphics[width=0.9\textwidth]{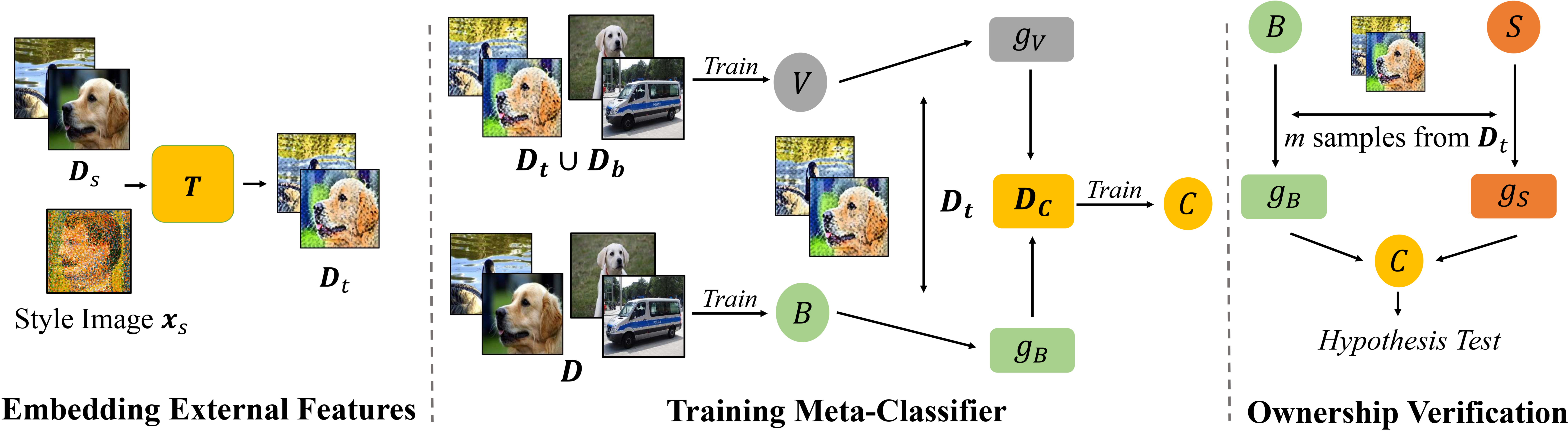}
    \vspace{-0.3em}
    \caption{The main pipeline of our method. In the first stage, defenders will modify some images via style transfer for embedding external features. In the second stage, defenders will train a meta-classifier to determine whether a suspicious model is stolen from the victim based on gradients. In the last stage, defenders will conduct ownership verification via hypothesis test.}
    \label{fig:pipeline}
    \vspace{-0.5em}
\end{figure*}

\vspace{0.3em}
\noindent \textbf{Results. }
As shown in Table \ref{table:models}, the ASR of the stolen model is only 3.84\%, which is significantly lower than that of the watermarked model. In other words, \emph{the defender-specified trigger no longer matches the hidden backdoors contained in the stolen model}. As such, backdoor-based model watermarking will fail to detect model stealing.

\begin{figure}[ht]
    \centering
    \subfigure[]{
		\label{fig:original}
		\includegraphics[width=0.143\textwidth]{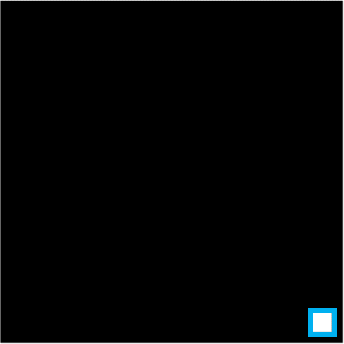}}
    \subfigure[]{
		\label{fig:before}
		\includegraphics[width=0.143\textwidth]{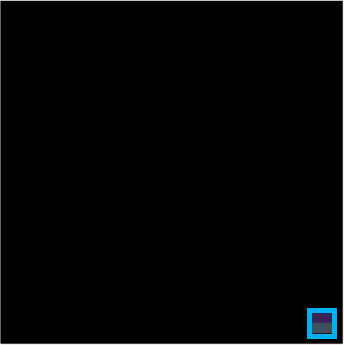}}
    \subfigure[]{
		\label{fig:after}
		\includegraphics[width=0.143\textwidth]{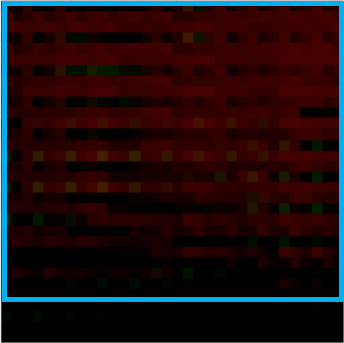}}
	\vspace{-0.7em}	
    \caption{The adopted trigger pattern and synthesized ones obtained from the watermarked and stolen model. The trigger areas are indicated in the blue box. \textbf{(a)} ground-truth trigger pattern; \textbf{(b)} pattern obtained from the watermarked model; \textbf{(c)} pattern obtained from the stolen model.} 
    \label{fig:backdoor-change}
    \vspace{-0.4em}
\end{figure}

To further understand the reason of this failure, we synthesize the potential trigger pattern of each model based on the targeted universal adversarial attack \cite{moosavi2017universal}. As shown in Figure \ref{fig:backdoor-change}, the pattern recovered from the watermarked model is similar to the ground-truth one, whereas the one recovered from the stolen model is completely different from the ground-truth pattern. These results explain why backdoor-based model watermarking has minor effects in defending against model stealing.

Moreover, \emph{backdoor-based model watermarking will introduce new security threats}, since it builds a stealthy latent connection between trigger pattern and target label. The adversary may use it to maliciously manipulate the predictions of deployed models. This problem will also hinder the utility of backdoor-based model watermarking in practice.

\section{The Proposed Method}
Based on the understandings in Section \ref{sec:limitation}, in this paper, we propose to embed \emph{external features} instead of inherent features for ownership verification. Specifically, as shown in Figure \ref{fig:pipeline}, our method consists of three main stages, including \textbf{(1)} embedding external features, \textbf{(2)} training an ownership meta-classifier, and \textbf{(3)} conducting ownership verification. Their technical details are in the following subsections.

\subsection{Threat Model}
Following the setting of existing works \cite{zhang2020passport,wang2021riga,liu2021watermarking}, we conduct the defense in a \emph{white-box} setting, where the defender has complete access to the suspicious model. However, the defender has no information about the stealing process. The goal of defenders is to accurately identify whether the suspicious model is stolen from a victim model, based on behaviors of the suspicious and victim model. 

One may argue that only black-box defenses are practical since the adversary may refuse to provide the suspicious model. However, white-box defenses are also practical. In our understanding, the real-world adoption of verification-based defenses (in a legal system) requires an official institute for the arbitration. Specifically, all commercial models should be registered here, through the unique identification ($e.g.$, MD5 code) of their model’s weights file. When this official institute is established, its staff should take responsibility for the verification process. For example, the staff can require the company to provide the model file with the same registered identification and then use our method (under the white-box setting) for the ownership verification. 

\subsection{Embedding External Features}
In this section, we describe how to embed external features into the victim model. Before we reach technical details, we first present the definition of inherent and external features.

\begin{definition}
A feature $f$ is called the inherent feature (of dataset $\mathcal{D}$) if and only if 
$\ 
\forall (\bm{x}, y) \in \mathcal{X}\times \mathcal{Y}, (\bm{x}, y) \in \mathcal{D} \Rightarrow (\bm{x}, y)\ \text{contains feature} f.
$
Similarly, f is called the external feature (of dataset $\mathcal{D}$) if and only if
$\ 
\forall (\bm{x}, y) \in \mathcal{X}\times \mathcal{Y}, (\bm{x}, y)\ \text{contains feature} \ f \Rightarrow (\bm{x}, y) \notin \mathcal{D}.
$
\end{definition}

Although external features are well defined, how to construct them is still difficult since the learning dynamic of DNNs remains unclear and the concept of features itself is complicated. However, at least we know that the \emph{image style} can serve as a feature for the learning of DNNs in image-related tasks, based on some recent studies \cite{geirhos2019imagenet,duan2020adversarial,cheng2021deep}. As such, we can use \emph{style transfer} \cite{johnson2016perceptual, huang2017arbitrary,chen2020optical} for embedding external features. People may also adopt other methods for the embedding. It will be discussed in our future work.

\begin{figure*}[ht]

    \centering
    \subfigure[]{
		\includegraphics[width=0.15\textwidth]{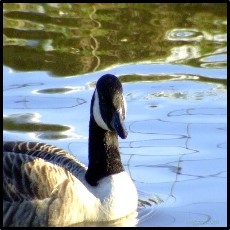}}
    \hspace{0.1em}
    \subfigure[]{
		\includegraphics[width=0.15\textwidth]{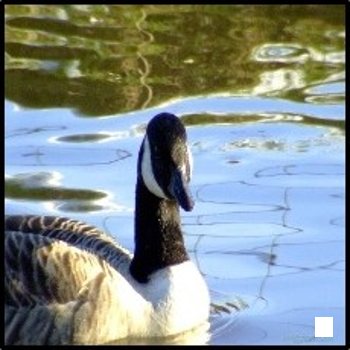}}
	\hspace{0.1em}
	\subfigure[]{
		\includegraphics[width=0.15\textwidth]{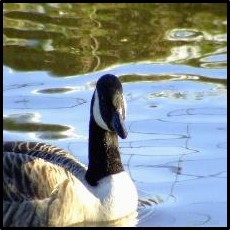}}
	\hspace{0.1em}
	\subfigure[]{
		\includegraphics[width=0.15\textwidth]{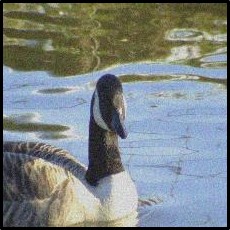}}
	\hspace{0.1em}
	\subfigure[]{
		\includegraphics[width=0.15\textwidth]{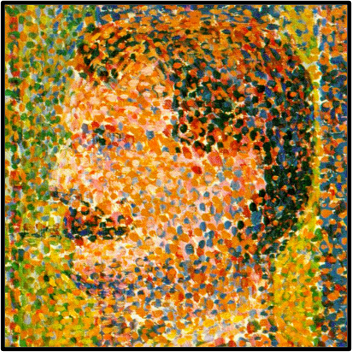}}
	\hspace{0.1em}
    \subfigure[]{
		\includegraphics[width=0.15\textwidth]{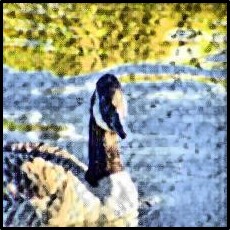}}
    \caption{Images involved in different defenses. \textbf{(a)} benign image; \textbf{(b)} poisoned image in BadNets; \textbf{(c)} poisoned image in Gradient Matching; \textbf{(d)} poisoned image in Entangled Watermarks; \textbf{(e)} style image; \textbf{(f)} transformed image. }
    \label{fig:images}
    \vspace{-0.4em}
\end{figure*}

Specifically, let $\mathcal{D} = \{ (\bm{x}_i, y_i) \}_{i=1}^{N}$ denotes the benign training set, $\bm{x}_s$ is a defender-specified \emph{style image}, and $T: \mathcal{X} \times \mathcal{X} \rightarrow \mathcal{X}$ is a (trained) style transformer. In this stage, the defender first randomly selects $\gamma \%$ (dubbed \emph{transformation rate}) samples ($i.e.$, $\mathcal{D}_s$) from $\mathcal{D}$ to generate their transformed version $\mathcal{D}_t = \{ (\bm{x}', y) |\bm{x}' = T(\bm{x}, \bm{x}_s), (\bm{x}, y) \in \mathcal{D}_s \}$. The external features (contained in $\bm{x}_s$) will be learned by the victim model $V_\theta$ during the training process via

\begin{equation}
\min_{\bm{\theta}} \sum_{(\bm{x}, y) \in \mathcal{D}_b \cup \mathcal{D}_t} \mathcal{L}(V_{\bm{\theta}}(\bm{x}), y),
\end{equation}
where $\mathcal{D}_b \triangleq \mathcal{D} \backslash \mathcal{D}_s$ and $\mathcal{L}(\cdot)$ is the loss function.

In this stage, how to select the style image is an important question. Intuitively, it should be significantly different from those contained in the original training set. In practice, defenders can simply adopt oil or sketch paintings as the style image since most of the images that need to be protected are natural images. We will further discuss it in Section \ref{sec:hyper}.

In particular, since we only poison a few samples and do not change their labels, the embedding of external features will not hinder the functionality of victim models or introduce new security risks ($e.g.$, hidden backdoors).

\comment{
\begin{figure}[ht]
    \centering
    \subfigure[Original image.]{
		\label{fig:original_images}
		\includegraphics[width=3.5cm]{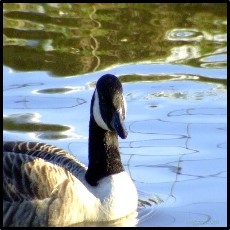}}
    \hspace{.15in}
    \subfigure[Style image.]{
		\label{fig:style}
		\includegraphics[width=3.5cm]{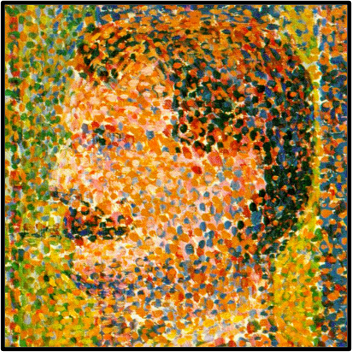}}
	\hspace{.15in}
    \subfigure[Embedded image.]{
		\label{fig:embedded}
		\includegraphics[width=3.5cm]{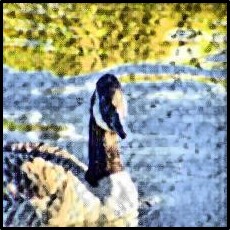}}
    \caption{The original image, style image, and the external feature embedded image using the style transfer technique \cite{johnson2016perceptual}}\label{fig:external-feature}
\end{figure}
}

\subsection{Training Ownership Meta-Classifier}
Since there is no explicit expression of the embedded external features and those features also have minor influences on the prediction, we need to train an additional binary meta-classifier to determine whether the suspicious model contains the knowledge of external features.

In this paper, we adopt the gradients of model weights as the input to train the meta-classifier $C_{\bm{w}}: \mathbb{R}^{|\bm{\theta}|} \rightarrow \{-1, +1\}$. Specifically, we assume that the victim model $V$ and the suspicious model $S$ have the same model structure. This assumption can be easily satisfied since the defender can retain a copy of the suspicious model on the training set of the deployed model as the victim model. Once the suspicious model is obtained, the defender will train its benign version ($i.e.$, the $B$) on the original training set $\mathcal{D}$. After that, we can obtain the training set $\mathcal{D}_c$ of meta-classifier $C$ via 
\begin{equation}
\begin{aligned}
   \mathcal{D}_c = & \left\{\left(g_V(\bm{x}'), +1\right)| (\bm{x}', y) \in \mathcal{D}_t \right\} \cup \\
   & \left\{\left(g_B(\bm{x}'), -1\right)| (\bm{x}', y) \in \mathcal{D}_t \right\}, 
\end{aligned}
\end{equation}
where $\text{sgn}(\cdot)$ is sign function \cite{sachs2012applied}, $g_{V}(\bm{x}') =  \text{sgn}( \nabla_{\bm{\theta}} \mathcal{L}(V(\bm{x}'), y))$, and $g_{B}(\bm{x}') =  \text{sgn}( \nabla_{\bm{\theta}} \mathcal{L}(B(\bm{x}'), y))$.

At the end, the meta-classifier $C_{\bm{w}}$ is trained by 
\begin{equation}
    \min_{\bm{w}} \sum_{(\bm{s}, t) \in \mathcal{D}_c} \mathcal{L}(C_{\bm{w}}(\bm{s}), t).
\end{equation}

In particular, we adopt its sign vector instead of the gradient itself to highlight the influence of its direction. We verify its effectiveness in \textbf{Appendix}.

\subsection{Ownership Verification with Hypothesis Test}
\label{sec:verification}
When the meta-classifier is trained, given a transformed image $\bm{x}'$ and its label $y$, the defender can examine the suspicious model simply by the result of $C(g_S(\bm{x}'))$, where $g_S(\bm{x}')= \text{sgn}( \nabla_{\bm{\theta}} \mathcal{L}(S(\bm{x}'), y))$. If $C(g_S(\bm{x}')) = 1$, the suspicious model is considered as stolen from the victim. However, it may be sharply affected by the randomness of selecting $\bm{x}'$. In this paper, we design a hypothesis test based method to increase the verification confidence, as follows:

\begin{definition}
Let $\bm{X}'$ denotes the variable of transformed images, while $\mu_{S}$ and $\mu_{B}$ indicates the posterior probability of the event $C(g_S(\bm{X}')) = 1$ and $C(g_B(\bm{X}')) = 1$, respectively. Given a null hypothesis $H_0: \mu_{S} \leq \mu_{B} \ (H_1: \mu_{S} > \mu_{B})$, we claim that the suspicious model $S$ is stolen from the victim if and only if the $H_0$ is rejected.
\end{definition}

In practice, we randomly sample $m$ different transformed images from $\mathcal{D}_t$ to conduct the pair-wise T-test \cite{hogg2005introduction} and calculate its p-value. When the p-value is smaller than the significance level $\alpha$, $H_0$ is rejected. Besides, we also calculate the \emph{confidence score} $\Delta \mu = \mu_{S} - \mu_{B}$ to represent the verification confidence. The larger the $\Delta \mu$, the more confident the verification.

\section{Experiments}
\subsection{Settings}
\label{sec:setting}
\textbf{Dataset and Model Selection. } 
We evaluate our defense on CIFAR-10 \cite{krizhevsky2009learning} and (a subset of) ImageNet \cite{deng2009imagenet} dataset. Following the settings of \cite{maini2021dataset}, we use the WideResNet \cite{zagoruyko2016wide} and ResNet \cite{he2016deep} as the victim model on CIFAR-10 and ImageNet, respectively. More detailed settings are in \textbf{Appendix}.

\begin{table*}[!ht]
  \centering
  \vspace{0.5em}
  \scalebox{0.86}{
  \begin{tabular}{llcccccccccccccc}  
  \toprule  
  \multicolumn{2}{c}{\multirow{2}*{Model Stealing}} 
  &\multicolumn{2}{c}{BadNets}& &\multicolumn{2}{c}{Gradient Matching}& &\multicolumn{2}{c}{Entangled Watermarks}& &\multicolumn{2}{c}{Dataset Inference}& & \multicolumn{2}{c}{Ours}\\
  \cline{3-4}\cline{6-7}\cline{9-10}\cline{12-13}\cline{15-16}
  & &$\Delta \mu$ & p-value & & $\Delta \mu$ & p-value & & $\Delta \mu$ & p-value & & $\Delta \mu$ & p-value & & $\Delta \mu$ & p-value\\
  \hline
  Victim & Source                                     &0.91       &$10^{-12}$ && 0.88      & $10^{-12}$ &&\textbf{0.99}&$\mathbf{10^{-35}}$ &&-&$10^{-4}$ & & 0.97         & $10^{-7}$\\ \hline
  $\mathcal{A}_{D}$ &Distillation                     &$-10^{-3}$ & 0.32      && $10^{-7}$ & 0.20       &&0.01         & 0.33               &&-&$10^{-4}$ & &\textbf{0.53} & $\mathbf{10^{-7}}$\\ 
  \multirow{2}*{$\mathcal{A}_{M}$}&Zero-shot          &$10^{-25}$ & 0.22      &&$10^{-24}$ & 0.22       &&$10^{-3}$    & $10^{-3}$          &&-&$10^{-2}$ & &\textbf{0.52} & $\mathbf{10^{-5}}$\\ 
  &Fine-tuning                                        &$10^{-23}$ & 0.28      &&$10^{-27}$ & 0.28       &&0.35& 0.01               &&-&$10^{-5}$ & &\textbf{0.50} & $\mathbf{10^{-6}}$\\ 
  \multirow{2}*{$\mathcal{A}_{Q}$}&Label-query        &$10^{-27}$ & 0.20      &&$10^{-30}$ & 0.34       &&$10^{-5}$    &0.62                &&-&$10^{-3}$ & &\textbf{0.52} & $\mathbf{10^{-4}}$\\ 
  &Logit-query                                        &$10^{-27}$ & 0.23      &&$10^{-23}$ & 0.33       &&$10^{-6}$    &0.64                &&-&$10^{-3}$ & &\textbf{0.54} & $\mathbf{10^{-4}}$ \\ \hline
  Benign&Independent                                  &$10^{-20}$ & 0.33      &&$10^{-12}$ & 0.99       &&$10^{-22}$   &0.68                &&-&\textbf{1.00} & &\textbf{0.00}  & \textbf{1.00} \\ 
  \bottomrule
  \end{tabular}
  }
\vspace{-0.8em}  
\caption{Results on CIFAR-10 dataset.}
  \label{table:main-cifar10} 
  \vspace{0.5em}
\end{table*}

\begin{table*}[!ht]
  \centering
  \scalebox{0.86}{
  \begin{tabular}{llcccccccccccccc}  
  \toprule  
  \multicolumn{2}{c}{\multirow{2}*{Model Stealing}} 
  &\multicolumn{2}{c}{BadNets}& &\multicolumn{2}{c}{Gradient Matching}& &\multicolumn{2}{c}{Entangled Watermarks}& &\multicolumn{2}{c}{Dataset Inference}& & \multicolumn{2}{c}{Ours}\\
  \cline{3-4}\cline{6-7}\cline{9-10}\cline{12-13}\cline{15-16}
  & &$\Delta \mu$ & p-value & & $\Delta \mu$ & p-value & & $\Delta \mu$ & p-value & & $\Delta \mu$ & p-value & & $\Delta \mu$ & p-value\\
  \hline
  Victim & Source                                     &0.87         &$10^{-10}$   &&0.77       &$10^{-10}$ &&\textbf{0.99} &$\mathbf{10^{-25}}$ &&-&$10^{-6}$&&0.90            &$10^{-5}$\\ \hline
  $\mathcal{A}_{D}$ &Distillation                     &$10^{-4}$    &0.43         &&$10^{-12}$ &0.43       &&$10^{-6}$     &0.19                &&-&$10^{-3}$&&\textbf{0.61}    &$\mathbf{10^{-5}}$\\
  \multirow{2}*{$\mathcal{A}_{M}$}&Zero-shot &$10^{-12}$   &0.33         &&$10^{-18}$ &0.43       &&$10^{-3}$     &0.46                &&-&$10^{-3}$&&\textbf{0.53}    &$\mathbf{10^{-4}}$\\
  &Fine-tuning                                        &$10^{-20}$   &0.20         &&$10^{-12}$ &0.47       &&0.46          &0.01                &&-&$10^{-4}$&&\textbf{0.60}    &$\mathbf{10^{-5}}$\\
  \multirow{2}*{$\mathcal{A}_{Q}$}&Label-query        &$10^{-23}$   &0.29         &&$10^{-22}$ &0.50       &&$10^{-7}$     &0.45                &&-&$\bm{10^{-3}}$&&\textbf{0.55}    &$\mathbf{10^{-3}}$\\
  &Logit-query                                        &$10^{-23}$   &0.38         &&$10^{-12}$ &0.22       &&$10^{-6}$     &0.36                &&-&$10^{-3}$&&\textbf{0.55}    &$\mathbf{10^{-4}}$\\ \hline
  Benign&Independent                                  &$10^{-24}$   &0.38         &&$10^{-23}$ &0.78       &&$\mathbf{10^{-30}}$    &0.55                &&-&0.98     &&$10^{-5}$&\textbf{0.99}\\
  \bottomrule
  \end{tabular}
  }
  \vspace{-0.8em}
\caption{Results on ImageNet dataset.}
  \label{table:main-imagenet}
  \vspace{0.1em}
\end{table*}

\vspace{0.3em}
\noindent \textbf{Settings for Model Stealing. }
Following the settings in \cite{maini2021dataset}, we conduct model stealing methods illustrated in Section \ref{sec:stealing-attack} to evaluate the effectiveness of defenses. Besides, we also provide the results of directly copying the victim model (dubbed `Source') and examining a suspicious model which is not stolen from the victim (dubbed `Independent') for reference. More detailed settings can be found in \textbf{Appendix}.

\vspace{0.3em}

\noindent \textbf{Defense Setup. } 
We compare our defense with dataset inference \cite{maini2021dataset} and model watermarking \cite{adi2018turning} with BadNets \cite{gu2019badnets}, gradient matching \cite{geiping2021witches}, and entangled watermarks \cite{jia2021entangled}. We poison 10\% training samples for all defenses. Besides, we adopt a white-square in the lower right corner as the trigger pattern for BadNets and adopt a oil paint as the style image for our defense. Other settings are the same as those used in their original paper. An example of images ($e.g.$, poisoned images and the style image) involved in different defenses is shown in Figure \ref{fig:images}.

\vspace{0.3em}
\noindent \textbf{Evaluation Metric.}
We use the confidence score $\Delta \mu$ and p-value for the evaluation metric. Following the settings adopted in \cite{maini2021dataset}, both $\Delta \mu$ and p-value are calculated based on the hypothesis test with 10 sampled images. In particular, except for the independent sources (which should not be regarded as stolen), \emph{the smaller the p-value and the larger the $\Delta \mu$, the better the defense}. Among all defenses, the best result is indicated in boldface.

\subsection{Main Results}
\label{Sec:mainResults}
As shown in Table \ref{table:main-cifar10}-\ref{table:main-imagenet}, our defense reaches the best performance in almost all cases. For example, the p-value of our method is three orders of magnitude smaller than that of the dataset inference and six orders of magnitude smaller than that of the model watermarking in defending against the distillation-based model stealing on CIFAR-10 dataset. The only exceptions appear when there is no model stealing. In these cases, entangled watermarks based model watermarking has some advantages. Nevertheless, our method can still easily make correct predictions in these cases. In particular, our defense method has minor adverse effects on the performance of victim models. For example, the accuracy of the model trained on benign CIFAR-10 and its transformed version is 91.99\% and 91.79\%, respectively. This is mainly because we do not change the label of transformed images and therefore the transformation can be treated as data augmentation, which is mostly harmless.

\begin{figure}[ht]
    \centering
    \includegraphics[width=0.45\textwidth]{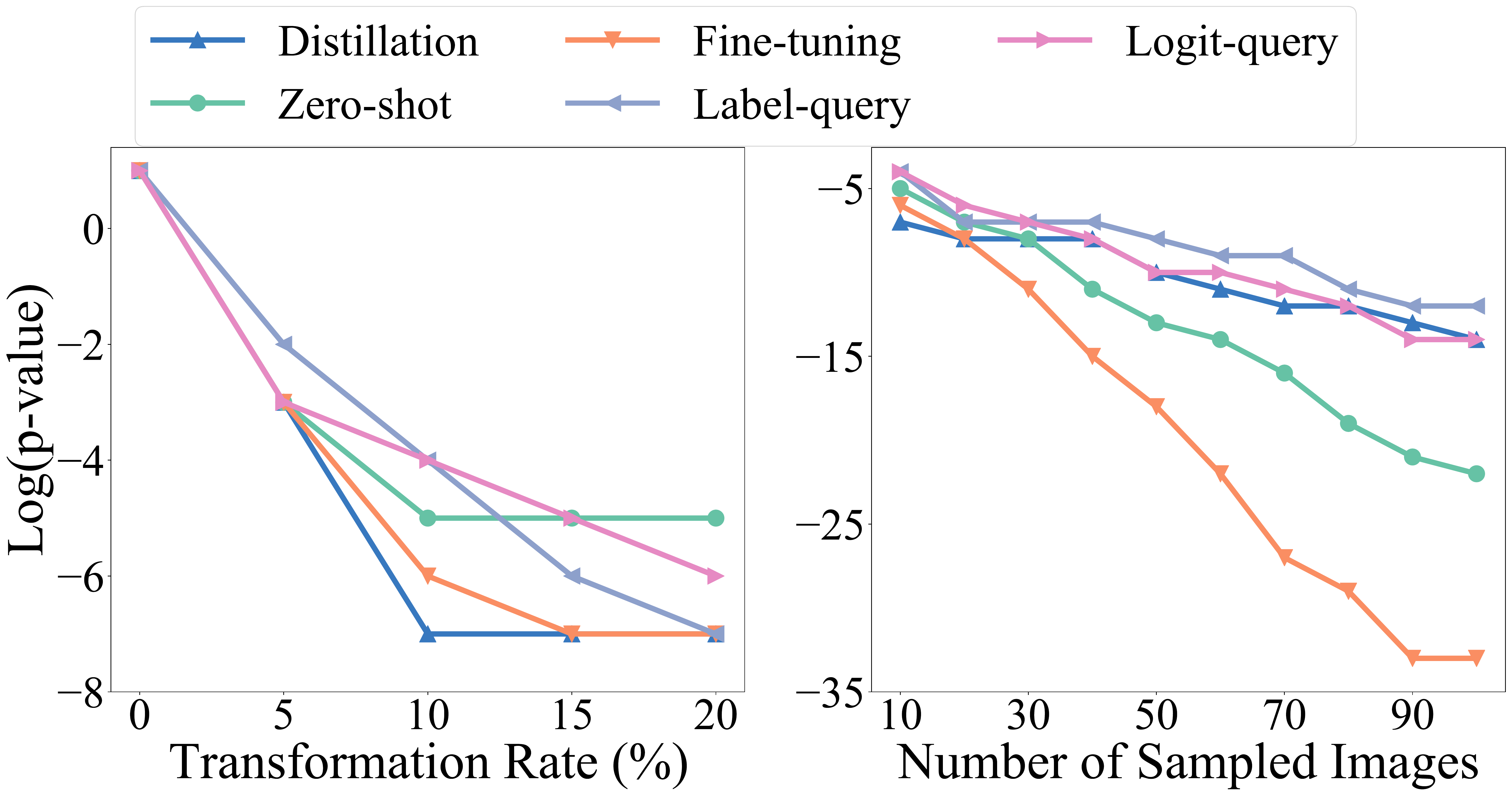}
    \vspace{-0.5em}
    \caption{Effects of the transformation rate (\%) and the number of sampled images.}
    \label{fig:hyper}
    \vspace{-0.3em}
\end{figure}

\begin{table*}[htbp]
  \centering
  \scalebox{1}{
  \begin{tabular}{llccccccccccc}  
  \toprule  
  \multicolumn{2}{c}{\multirow{2}*{Model Stealing}} 
  & \multicolumn{2}{c}{Pattern (a)}& &\multicolumn{2}{c}{Pattern (b)}& &\multicolumn{2}{c}{Pattern (c)}& &\multicolumn{2}{c}{Pattern (d)}\\
  \cline{3-4}\cline{6-7}\cline{9-10}\cline{12-13}
  & &$\Delta \mu$ & p-value& &$\Delta \mu$ & p-value& &$\Delta \mu$ & p-value& &$\Delta \mu$ & p-value\\
  \hline
  Victim & Source                                     & 0.98 & $10^{-7}$ &&0.97 & $10^{-7}$ && 0.98 & $10^{-10}$ && 0.98    & $10^{-12}$ \\ \hline 
  \multirow{1}*{$\mathcal{A}_{D}$} &Distillation      & 0.68 & $10^{-7}$ &&0.53 & $10^{-7}$ && 0.72 & $10^{-8}$  && 0.63    & $10^{-7}$\\ 
  \multirow{2}*{$\mathcal{A}_{M}$}&Zero-shot          & 0.61 & $10^{-5}$ &&0.52 & $10^{-5}$ && 0.74 & $10^{-8}$  && 0.67    & $10^{-7}$  \\ 
  &Fine-tuning                                        & 0.46 & $10^{-5}$ &&0.50 & $10^{-6}$ && 0.21 & $10^{-7}$  && 0.50    & $10^{-9}$\\ 
  \multirow{2}*{$\mathcal{A}_{Q}$}&Label-query        & 0.64 & $10^{-5}$ &&0.52 & $10^{-4}$ && 0.68 & $10^{-8}$  && 0.68    & $10^{-7}$\\ 
  &Logit-query                                        & 0.65 & $10^{-4}$ &&0.54 & $10^{-4}$ && 0.62 & $10^{-6}$  && 0.73    & $10^{-7}$\\ \hline
  Benign&Independent                                  & 0.00   & 1.00      &&0.00  & 1.00       && 0.00  & 1.00        &&$10^{-9}$& 0.99  \\
  \bottomrule
  \end{tabular}
  }
  \vspace{-0.8em}
\caption{The effectiveness of our defense with different style images on CIFAR-10 dataset.}
  \label{table:dis-another}  
  \vspace{0.3em}
\end{table*}

\subsection{Discussion}
\label{sec:hyper}
In this section, we discuss the effects of hyper-parameters and components involved in our method. Unless otherwise specified, all settings are the same as those in Section \ref{Sec:mainResults}.

\vspace{0.3em}
\noindent \textbf{Effects of Transformation Rate. } 
The larger the transformation rate $\gamma$, the more training samples are transformed during the training process of the victim model. As we expected, the p-value decrease with the increase of $\gamma$ in defending all stealing methods (as shown in Figure \ref{fig:hyper}). Note that the increase of $\gamma$ may also lead to the accuracy decrease of victim models. Defenders should specify this hyper-parameter based on their specific requirements in practice.

\vspace{0.3em}
\noindent \textbf{Effects of the Number of Sampled Images. }
Recall that our method needs to specify the number of sampled (transformed) images ($i.e.$, the $m$) adopted in the hypothesis-based ownership verification. In general, the larger the $m$, the less the adverse effects of the randomness involved in this process and therefore the more confident the verification. 
This is probably the reason why the p-value also decreases with the increase of $m$, as shown in Figure \ref{fig:hyper}.

\begin{table}[htbp]
  \centering
  \scalebox{0.79}{
  \begin{tabular}{lccccc}  
  \toprule  
   & \multicolumn{2}{c}{Style Transfer}& &\multicolumn{2}{c}{Meta-classifier} \\
  \cline{2-3}\cline{5-6}
  & \tabincell{c}{Patch-based\\Variant} & Ours& & BadNets &\tabincell{c}{BadNets +\\Meta-classifier}\\\hline
  Distillation & 0.17 & $\mathbf{10^{-7}}$ &&0.32 & $\mathbf{10^{-3}}$\\ 
  Zero-shot & 0.01 & $\mathbf{10^{-5}}$ &&0.22 & $\mathbf{10^{-61}}$\\ 
  Fine-tuning & $10^{-3}$ & $\mathbf{10^{-6}}$ &&0.28 & $\mathbf{10^{-5}}$\\ 
  Label-query & $10^{-3}$ & $\mathbf{10^{-4}}$ &&0.20 & $\mathbf{10^{-50}}$\\ 
  Logit-query & $10^{-3}$ & $\mathbf{10^{-4}}$ &&0.23 & $\mathbf{10^{-3}}$\\
  \bottomrule
  \end{tabular}
  }
  \vspace{-0.8em}
\caption{The effectiveness (p-value) of style transfer and meta-classifier on CIFAR-10 dataset.}
  \vspace{0.2em}
  \label{table:effect2}   
\end{table}

\begin{figure}[t]
    \centering
    \subfigure[]{
		\includegraphics[width=0.105\textwidth]{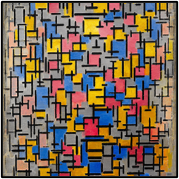}}
    \subfigure[]{
		\includegraphics[width=0.105\textwidth]{./fig/style.png}}
	\subfigure[]{
		\includegraphics[width=0.105\textwidth]{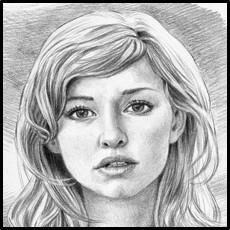}}
	\subfigure[]{
		\includegraphics[width=0.105\textwidth]{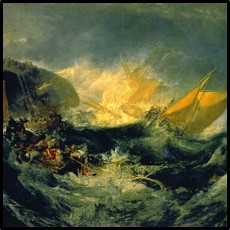}}
	\vspace{-0.8em}
    \caption{Style images adopted for the evaluation.}
    \vspace{-0.2em}
    \label{fig:styimg}
\end{figure}

\vspace{0.3em}
\noindent \textbf{Effects of Style Image. }
In this part, we examine whether the proposed defense is still effective if we adopt other style images (as shown in Figure \ref{fig:styimg}). As shown in Table \ref{table:dis-another}, the p-value is significantly smaller than 0.01 in all cases. In other words, our method remains effective in defending against different stealing methods when different style images are used, although there will be some fluctuations in the results. We will further explore how to optimize the selection of style images in our future work.

\vspace{0.3em}
\noindent \textbf{Effectiveness of Style Transfer. }
To verify that the style watermark transfers better during the stealing process, we compare our method with its variant which uses the white-square patch (adopted in BadNets) to generate transformed images. As shown in Table \ref{table:effect2}, our method is significantly better than its patch-based variant. It is probably because DNNs are easier to learn the texture information \cite{geirhos2019imagenet} and the style watermark is bigger than the patch one. This phenomenon partly explains why our method works well.

\vspace{0.3em}
\noindent \textbf{Effectiveness of Meta-Classifier. }
To verify that the meta-classifier is also useful, we compare the BadNets-based model watermarking with its extension which also uses the meta-classifier (adopted in our method) for ownership verification. In this case, the victim model is the backdoored one and the transformed image is the one containing backdoor triggers. As shown in Table \ref{table:effect2}, adopting meta-classier significantly decrease the p-value in all cases, which verifies the effectiveness of the meta-classifier. These results also partly explains the effectiveness of our method.

\begin{table}[!t]
  \centering

    \scalebox{0.83}{
  \begin{tabular}{c|ccc}  
  \toprule
 Attack Stage $\rightarrow$ & Source  &First Stage & Second Stage\\
    \hline
    Attack Method $\rightarrow$ & None    & Zero-shot &Zero-shot\\
    Model Structure $\rightarrow$ &WRN-28-10 &WRN-28-10           &WRN-28-10 \\
    p-value $\rightarrow$      &$10^{-7}$ &$10^{-5}$           &$10^{-4}$ \\
    \hline \hline
    Attack Method $\rightarrow$  & None    & Logit-query        &Zero-shot\\
    Model Structure $\rightarrow$ &WRN-28-10 &WRN-16-1            &VGG-19\\
    p-value $\rightarrow$             &$10^{-7}$ &$10^{-4}$           &0.01 \\
  \bottomrule
  \end{tabular}
  }
  \vspace{-0.8em}
\caption{Results in defending against multi-stage stealing.}
  \label{table:dis-multi}  
\end{table}

\vspace{0.3em}
\subsection{Defending against Multi-Stage Model Stealing}
\label{sec:dis}
\vspace{0.2em}
In previous experiments, the stolen model is obtained by a single stealing attack. In this section, we explore whether our method is still effective if there are multiple stealing stages.

\vspace{0.3em}
\noindent \textbf{Settings. }
We discuss two types of multi-stage stealing on the CIFAR-10 dataset, including \textbf{(1)} stealing with the same attack and model structure and \textbf{(2)} stealing with different attacks and model structures. In general, the first one is the easiest multi-stage attack while the second one is the hardest. Other settings are the same as those used in Section \ref{Sec:mainResults}.

\vspace{0.3em}
\noindent \textbf{Results. }
As shown in Table \ref{table:dis-multi}, the p-value $\leq 0.01$ in all cases, $i.e.$, our method can successfully identify the existence of model stealing, even after multiple stealing stages. As we expected, the p-value in defending the second multi-stage attack is significantly larger than that of the first one indicating that the second task is harder. We will discuss how to better defend the second type of attack in our future work.

\vspace{0.5em}
\section{Conclusion}
\vspace{0.4em}
In this paper, we formulated the defense of model stealing as verifying whether a suspicious model contains the knowledge of defender-specified external features. Specifically, we embedded external features by modifying a few training samples with style transfer. This approach was inspired by the understanding that the stolen models should contain the knowledge of features learned by the victim model. We evaluated our defense on both CIFAR-10 and ImageNet datasets, which verified that our method can defend against various types of model stealing simultaneously while preserving high accuracy in predicting benign samples. 

\newpage

\section*{Acknowledgments}
This work is supported in part by the National Key R\&D Program of China under Grant 2019YFB1406500, the National Natural Science Foundation of China (62171248, U1736219, U1936210), the Guangdong Province Key Area R\&D Program under Grant 2018B010113001, the R\&D Program of Shenzhen (JCYJ20180508152204044), and the PCNL KEY project (PCL2021A07).

\bibliography{aaai22}

\newpage

\begin{center}
   \LARGE \textbf{Appendix} 
\end{center}

\begin{appendices}
\section{Detailed Settings for Revisiting Defenses}

\subsection{The Limitation of Dataset Inference}
\label{sec:lim_dataset}
\textbf{Training Setups. } 
As described in Section 3.1 of the main manuscript, we train the VGG-19 on $D_l$ (dubbed VGG-${D_l}$), the ResNet18 on $D_r$ (dubbed ResNet-$D_r$), and another VGG-19 on the noisy dataset ${D'}_l$ (dubbed VGG-${D'}_l$), respectively. We train these three models for 200 epochs with the SGD optimizer and an initial learning rate of 0.1, momentum of 0.9, weight decay of 5 $\times 10^{-4}$, and batch size of 128. We decay the learning rate with the cosine decay schedule \cite{loshchilov2016sgdr} without a restart. 
We also use data augmentation techniques including random crop and resize (with random flip).
The training is conducted based on the open-source code\footnote{\url{https://github.com/verazuo/badnets-pytorch}}.

\vspace{0.3em}
\noindent \textbf{Verification Setups. } 
Following the setting of dataset inference \cite{maini2021dataset}, we extract feature embeddings from VGG-${D_l}$ with ${D_l}$, ResNet-$D_r$ with ${D_r}$ and VGG-${D'}_l$ with ${D'}_l$ respectively.
And then we train a binary meta-classifier for 1,000 epochs based on the feature embeddings of ResNet-$D_r$ with SGD optimizer and a learning rate of 0.1.
This meta-classifier is used to verify if VGG-${D_l}$ and VGG-${D'}_l$ is stolen from ResNet-$D_r$.
Another binary meta-classifier is trained with the same setting based on the feature embeddings of VGG-${D_l}$ to verify if ResNet-$D_r$ is stolen from VGG-${D_l}$. The verification is conducted based on the open-sourced code of dataset inference\footnote{\url{https://github.com/cleverhans-lab/dataset-inference}}. Besides, both training and verification are implemented on a single GeForce GTX 1080 Ti GPU.

\subsection{The Limitation of Model Watermarking}
\label{sec:lim_backdoor}

\vspace{0.3em}
\noindent \textbf{Training Setups. } 
We adopt a white-square as the trigger pattern and train a BadNets-based watermarked model with target label $y_t = 2$ by modifying $10\%$ training samples on CIFAR-10 dataset \cite{krizhevsky2009learning}. We use the WideResNet \cite{zagoruyko2016wide} with depth 28 and widening factor of 10 (WRN-28-10) as the model architecture. 
The training is performed on a single GeForce GTX 1080 Ti GPU and other training settings are the same as those described in Section \ref{sec:lim_dataset}.

\vspace{0.3em}
\noindent \textbf{Stealing Setups. } 
We perform the zero-shot learning based data-free distillation \cite{fang2019data} to obtain a stolen copy of each model. The attack is implemented based on the open-source code\footnote{\url{https://github.com/VainF/Data-Free-Adversarial-Distillation}}.
Specifically, the stealing attack is performed for 200 epochs with SGD optimizer and a learning rate of 0.1, momentum of 0.9, weight decay of $5\times 10^{-4}$, and batch size of 256. The stolen model has the same model structure as the victim model. The attack is performed on two GeForce GTX 1080 Ti GPUs.

\vspace{0.3em}
\noindent \textbf{Settings for Universal Adversarial Attack. } %
We use the projected gradient descent \cite{madry2017towards} method to generate adversarial perturbations within the ${\ell}^{\infty}$-ball where the maximum distortion size $\epsilon= 32$.
The adversarial perturbations are initialized as the trigger pattern used for watermarking. We implement the universal adversarial attack for 40 iterations on the backdoored model and the stolen model.

\section{Detailed Settings for Main Experiments}
\subsection{Dataset Description}
\label{sec:dataset}
We evaluate our method on both CIFAR-10 \cite{krizhevsky2009learning} and ImageNet \cite{deng2009imagenet} dataset. 
CIFAR-10 contains 60,000 images (with size 3x32x32) in 10 classes, including 50,000 training images and 10,000 testing images. ImageNet is a large-scale dataset and we use its subset containing 20 random classes to evaluate our method. Each class of the subset contains 500 images (with size 3x224x224) for training and 50 images for testing. 

\subsection{Detailed Settings for Training}
\label{sec:train_setup}
Following the settings used in \cite{maini2021dataset}, we use the WRN-16-1 \cite{zagoruyko2016wide} and ResNet-18 \cite{he2016deep} as the victim and benign model on CIFAR-10 and ImageNet, respectively. For the CIFAR-10 dataset, the training is conducted based on the open-source code$\footnote{\url{https://github.com/kuangliu/pytorch-cifar}}$. Specifically, both victim model and benign model are trained for 200 epochs with SGD optimizer and an initial learning rate of 0.1, momentum of 0.9, weight decay of 5 $\times 10^{-4}$, and batch size of 128. We decay the learning rate with the cosine decay schedule \cite{loshchilov2016sgdr} without a restart. We also use data augmentation techniques including random crop and resize (with random flip). For the ImageNet dataset, both the victim model and benign model are trained for 200 epochs with SGD optimizer and an initial learning rate of 0.001, momentum of 0.9, weight decay of $1 \times 10^{-4}$, and batch size of 32. The learning rate is decreased by a factor of 10 at epoch 150. All training processes are performed on a single GeForce GTX 1080 Ti GPU.

\subsection{Detailed Settings for Model Stealing}
\label{sec:attack_setup}
Following the setting adopted in dataset inference \cite{maini2021dataset}, we conduct model stealing with three different adversary's permission levels, including \textbf{(1)} dataset-accessible attacks, \textbf{(2)} model-accessible attacks, and \textbf{(3)} query-only attacks. Besides, we also provide the results of directly copying the victim model (dubbed `Source') and examining a suspicious model which is not stolen from the victim (dubbed `Independent') for reference.

Except for distillation-based methods ($i.e.$, Distillation and Zero-Shot), all methods are performed on a single GeForce GTX 1080 Ti GPU. We adopt two GeForce GTX 1080 Ti GPUs for distillation-based methods. Besides, except for fine-tuning, we adopt WRN-16-1 and ResNet-18 as the stolen model on CIFAR-10 and ImageNet, respectively. More detailed settings are as follows:

\vspace{0.3em}
\noindent \textbf{Distillation. }
We implement the model distillation \cite{hinton2015distilling} based on its open-sourced code$\footnote{\url{https://github.com/thaonguyen19/ModelDistillation-PyTorch}}$. 
Specifically, the stolen model is trained with SGD optimizer and an initial learning rate of 0.1, momentum of 0.9, weight decay of $10^{-4}$. We use ReduceLROnPlateau as the learning rate scheduler. Specifically, we set the mode as `min', factor as 0.1, and patience as 10.

\vspace{0.3em}
\noindent \textbf{Zero-shot. }
We perform zero-shot learning based data-free distillation \cite{fang2019data} to obtain the stolen model. All settings are the same as those described in Section \ref{sec:lim_backdoor}.

\vspace{0.3em}
\noindent \textbf{Fine-tuning. }
In this setting, attackers obtain stolen models by fine-tuning victim models on different datasets. Following the settings in \cite{maini2021dataset}, we randomly select 500,000 samples from the original TinyImages \cite{birhane2021large} as the substitute data to fine-tune the victim model for experiments on CIFAR-10. For the ImageNet experiments, we randomly choose samples with another 20 classes from the original ImageNet as the substitute data. We fine-tune the victim model for 5 epochs with settings described in Section \ref{sec:train_setup}.

\vspace{0.4em}
\noindent \textbf{Label-query. }
Following the settings used in dataset inference \cite{maini2021dataset}, we train the stolen model for 20 epochs with a substitute dataset labeled by the victim model.
Other settings are the same as those described in Section \ref{sec:train_setup}.

\vspace{0.3em}
\noindent \textbf{Logit-query. }
In this setting, attackers train the stolen model by minimizing the KL-divergence between its outputs ($i.e.$, logits) and those of the victim model. Following the settings used in dataset inference \cite{maini2021dataset}, we train the stolen model for 20 epochs with the same settings stated in Section \ref{sec:train_setup}.

\section{Effectiveness of the Sign Function}
In this section, we verify the effectiveness of adopting the sign vector of gradients instead of gradients themselves in the training of our meta-classifier.

\vspace{0.3em}
\noindent \textbf{Settings. }
We conduct the experiments on CIFAR-10 and ImageNet dataset. Except for the discussed component, other settings are the same as those used in Section \ref{sec:dataset}-\ref{sec:attack_setup}.

\vspace{0.3em}
\noindent \textbf{Results. }
As shown in Table \ref{table:signeffects}, adopting the sign of gradients is significantly better than adopting gradients directly. This is probably because the `direction' of gradients contains more information compared with their `magnitude'. We will further explore it in our future work.

\begin{table*}[ht]
\centering
\scalebox{1}{
\begin{tabular}{c|cc|cc|cccc}
\toprule
Dataset$\rightarrow$ & \multicolumn{4}{c|}{CIFAR-10} & \multicolumn{4}{c}{ImageNet}                                                                       \\ \hline
\multirow{2}{*}{Model Stealing$\downarrow$} & \multicolumn{2}{c|}{Gradient} & \multicolumn{2}{c|}{Sign of Gradient (Ours)} & \multicolumn{2}{c|}{Gradient} & \multicolumn{2}{c}{Sign of Gradient (Ours)} \\ \cline{2-9} 
& $\Delta \mu$ & p-value & $\Delta \mu$ & p-value & $\Delta \mu$ & \multicolumn{1}{c|}{p-value}& $\Delta \mu$ & p-value \\ \hline
Source       & 0.44 & $10^{-5}$ & $\bm{0.97}$ & $\bm{10^{-7}}$ &0.15      & \multicolumn{1}{c|}{$10^{-4}$} & $\bm{0.90}$ & $\bm{10^{-5}}$ \\ \hline
Distillation & 0.27 & 0.01      & $\bm{0.53}$ & $\bm{10^{-7}}$ &0.15      & \multicolumn{1}{c|}{$10^{-4}$} & $\bm{0.61}$ & $\bm{10^{-5}}$ \\
Zero-shot    & 0.03 & $10^{-3}$ & $\bm{0.52}$ & $\bm{10^{-5}}$ &0.12      & \multicolumn{1}{c|}{$10^{-3}$} & $\bm{0.53}$ & $\bm{10^{-4}}$ \\
Fine-tuning  & 0.04 & $10^{-5}$ & $\bm{0.50}$ & $\bm{10^{-6}}$ &0.13      & \multicolumn{1}{c|}{$10^{-3}$} & $\bm{0.60}$ & $\bm{10^{-5}}$ \\
Label-query  & 0.08 & $10^{-3}$ & $\bm{0.52}$ & $\bm{10^{-4}}$ &0.13      & \multicolumn{1}{c|}{$10^{-3}$} & $\bm{0.55}$ & $\bm{10^{-3}}$ \\
Logit-query  & 0.07 & $10^{-5}$ & $\bm{0.54}$ & $\bm{10^{-4}}$ &0.12      & \multicolumn{1}{c|}{$10^{-3}$} & $\bm{0.55}$ & $\bm{10^{-4}}$ \\ \hline
Independent & $\bm{0.00}$ & $\bm{1.00}$ & $\bm{0.00}$ & $\bm{1.00}$ &$\bm{10^{-10}}$ & \multicolumn{1}{c|}{$\bm{0.99}$} & $10^{-5}$ & $\bm{0.99}$ \\ \bottomrule
\end{tabular}
}
\vspace{-0.8em}
\caption{The performance of our meta-classifier trained with different features.}
\label{table:signeffects} 
\end{table*}

\begin{table*}[!t]
\centering
\scalebox{1}{
\begin{tabular}{c|cc|cc|cccc}
\toprule
Dataset$\rightarrow$ & \multicolumn{4}{c|}{CIFAR-10} & \multicolumn{4}{c}{ImageNet}                                                                       \\ \hline
\multirow{2}{*}{Model Stealing$\downarrow$} & \multicolumn{2}{c|}{Training Set} & \multicolumn{2}{c|}{Testing Set} & \multicolumn{2}{c|}{Training Set} & \multicolumn{2}{c}{Testing Set} \\ \cline{2-9} 
& $\Delta \mu$ & p-value & $\Delta \mu$ & p-value & $\Delta \mu$ & \multicolumn{1}{c|}{p-value}& $\Delta \mu$ & p-value \\ \hline
Source      & 0.97 & $10^{-7}$ & 0.96 & $10^{-7}$ & 0.90 & \multicolumn{1}{c|}{$10^{-5}$} & 0.93      & $10^{-7}$ \\ \hline
Distillation& 0.53 & $10^{-7}$ & 0.53 & $10^{-5}$ & 0.61 & \multicolumn{1}{c|}{$10^{-5}$} & 0.42      & $10^{-5}$ \\
Zero-shot   & 0.52 & $10^{-5}$ & 0.53 & $10^{-5}$ & 0.53 & \multicolumn{1}{c|}{$10^{-4}$} & 0.34      & $10^{-3}$ \\
Fine-tuning & 0.50 & $10^{-6}$ & 0.47 & $10^{-6}$ & 0.60 & \multicolumn{1}{c|}{$10^{-5}$} & 0.72      & $10^{-5}$ \\
Label-query & 0.52 & $10^{-4}$ & 0.52 & $10^{-4}$ & 0.55 & \multicolumn{1}{c|}{$10^{-3}$} & 0.40      & $10^{-3}$ \\
Logit-query & 0.54 & $10^{-4}$ & 0.53 & $10^{-4}$ & 0.55 & \multicolumn{1}{c|}{$10^{-4}$} & 0.48      & $10^{-4}$ \\ \hline
Independent & 0.00 & 1.00      & 0.00 & 1.00      & $10^{-5}$ & \multicolumn{1}{c|}{0.99} & $10^{-9}$ & 0.99 \\ \bottomrule
\end{tabular}
}
\vspace{-0.8em}
\caption{Results of ownership verification with transformed samples generated by samples in training and testing dataset.}
\label{table:membership}
\end{table*}

\section{Relations with Related Works}
In this section, we discuss the similarities and differences between our defense and related methods.

\subsection{Relations with Property Inference Attacks}
Our defense is inspired by the property inference attacks \cite{ganju2018property,wu2020joint,luo2021feature}, which intend to identify whether a particular property ($i.e.$, attribute or feature) is contained in the training set of given DNNs. Similar to these attacks, our defense also examines whether a suspicious model contains the knowledge of predefined external features. However, different from inference attacks whose targets are attributes directly contained in the training set, the concept of features is more complicated in our method. In our defense, the feature is defined as a complex combination of attributes ($i.e.$, pixel values).

\subsection{Relations with Membership Inference}
Membership inference attacks \cite{shokri2017membership,leino2020stolen, hui2021practical} aim to identify whether a particular sample is used to train given DNNs. Similar to these attacks, our method also adopts some training samples for ownership verification. However, our defense intend to analyze whether the suspicious model contains the knowledge of external features rather than whether the model is trained on those transformed samples. To verify it, we design additional experiments, as follows:

\vspace{0.3em}
\noindent \textbf{Settings. }
We compare our method with its variant which adopts testing images to generate transformed ones used in ownership verification. Except for this, other settings are the same as those used in Section 5.2 of the main manuscript.

\vspace{0.3em}
\noindent \textbf{Results. }
As shown in Table \ref{table:membership}, adopting testing images has similar effects to that of using training images in generating transformed images used for ownership verification. This phenomenon verifies that our defense is fundamentally different from membership inference attacks.

\subsection{Relations with Backdoor Attacks}
Similar to that of (poisoning-based) backdoor attacks \cite{li2020backdoor,zhao2020clean,zhai2021backdoor}, our defense embeds predefined behaviors into DNNs through modifying some training samples. However, different from that of backdoor attacks, our method neither changes the label of poisoned samples (to the target label) nor only selects training samples with the specific category ($i.e.$, target label) for poisoning. As such, the embedding of external features adopted in our defense will not introduce hidden backdoors into the trained deployed model. 

\end{appendices}

\end{document}